# Line tensions, correlation lengths, and critical exponents in lipid membranes near critical points

Aurelia R. Honerkamp-Smith[1], Pietro Cicuta[2], Marcus D. Collins[1], Sarah L. Veatch[3], Marcel den Nijs[4], M. Schick[4], and Sarah L. Keller[1,4]*

[1.] Dept. of Chemistry, University of Washington, Seattle WA 98195, USA
[2.] Cavendish Laboratory, University of Cambridge, Cambridge, CB3 0HE, UK
[3.] Dept. of Chemistry, Cornell University, Ithaca, NY, 14853, USA
[4.] Dept. of Physics, University of Washington, Seattle WA 98195, USA
* To whom correspondence should be addressed.

**ABSTRACT**

Membranes containing a wide variety of ternary mixtures of high chain-melting temperature lipids, low chain-melting temperature lipids, and cholesterol undergo lateral phase separation into coexisting liquid phases at a miscibility transition. When membranes are prepared from a ternary lipid mixture at a critical composition, they pass through a miscibility critical point at the transition temperature. Since the critical temperature is typically on the order of room temperature, membranes provide an unusual opportunity in which to perform a quantitative study of biophysical systems that exhibit critical phenomena in the two-dimensional Ising universality class. As a critical point is approached from either high or low temperature, the scale of fluctuations in lipid composition, set by the correlation length, diverges. In addition, as a critical point is approached from low temperature, the line tension between coexisting phases decreases to zero. Here we quantitatively evaluate the temperature dependence of line tension between liquid domains and of fluctuation correlation lengths in lipid membranes in order to extract a critical exponent, $\nu$. We obtain $\nu = 1.2 \pm 0.2$, consistent with the Ising model prediction $\nu = 1$. We also evaluate the probability distributions of pixel intensities in fluorescence images of membranes. From the temperature dependence of these distributions above the critical temperature, we extract an independent critical exponent of $\beta = 0.124 \pm 0.03$, which is consistent with the Ising prediction of $\beta = 1/8$.

**Symbols and Abbreviations:**
$T_c$ - critical temperature, $\lambda$ - line tension, $\xi$ - correlation length, $k_B$ - Boltzmann's constant, $\nu$ - critical exponent for the correlation length, $\beta$ - critical exponent for the order parameter

## INTRODUCTION

Simple lipid bilayers undergo complex physical behavior. Consider a bilayer membrane composed of three components: dipalmitoylphosphatidylcholine (DPPC), a lipid with a high chain-melting temperature (41°C (1)), diphytanoylphosphatidylcholine (diPhyPC), a lipid with a low chain-melting temperature (less than -120°C (2)), and cholesterol. At high temperatures, all lipids in the bilayer mix uniformly in a single liquid phase. As the temperature is lowered, a transition occurs and the lipids separate laterally into coexisting liquid phases (3). As a result, shortly after the membrane passes through the phase transition, it contains domains of one phase whose lipid composition differs from that of the surrounding phase. All lipids in the membrane diffuse with rates characteristic of liquids (4-6). Lipids in each phase are distinguished by the degree of order in the acyl chains, which is higher in the liquid-ordered ($L_o$) phase than in the liquid-disordered ($L_d$) phase (7). Fluorescently-labeled lipids partition differently into the $L_o$ and $L_d$ phases of vesicle membranes, which permits straightforward imaging of the different phases, as in figure 1 and in the movie included in a supplementary file.

Miscibility transition temperatures vary with the membrane's total lipid composition. In a three-dimensional plot with compositions in the *x-y* plane and transition temperatures in the *z* direction, the phase boundary resembles the surface of a hill. Its crest is an upper critical point. At lower temperatures, a line of critical points forms a path up one side of the hill, and down the other (8). At each composition corresponding to a critical point, the phase transition occurs at a particular critical temperature, $T_c$. Miscibility transitions that occur at a critical point are identifiable by specific features described below.

We observe clear experimental signatures of critical behavior in a membrane. Both above and below the critical temperature $T_c$, fluctuations in lipid composition appear over all distances up to a characteristic correlation length, $\xi$. As $T_c$ is approached, the correlation length diverges according to a power law $\xi \sim |T - T_c|^{-\nu}$, where $\nu$ is the critical exponent. We observe an additional experimental signature of critical behavior at temperatures below $T_c$. Domain boundaries between the two liquid phases fluctuate (3; 8-12). As $T_c$ is approached from below, the compositions of the two membrane phases become increasingly similar, until they are identical at the critical point. The difference in lipid compositions of the two phases, the order parameter *m*, vanishes as $m \sim (T_c - T)^\beta$. Consequently, the line tension, $\lambda$, between domain boundaries in two-dimensional systems decreases and becomes zero at the critical point such that $\lambda \sim (T_c - T)^\nu$. When the interface energy becomes as low as the thermal energy, fluctuations in domain boundaries become visible optically (see figure 1). Low line tensions (less than 1pN) (13-15), and significant composition fluctuations (16) have been observed previously in lipid membranes near critical points. We also observe the divergence of composition fluctuations above the critical point to determine the exponent *β*.

We employ fluorescence microscopy and image processing to systematically measure the dependence of membrane line tensions and correlation lengths on temperature in order to extract the critical exponents $\nu$ and $\beta$. These two independent exponents are characteristic of the universality class of the transition. We verify our expectation that critical behavior in lipid bilayers is in the universality class of the two-dimensional Ising model, for which $\nu$ and $\beta$ are known to be 1 and 1/8, respectively.

The Ising model is a simple theoretical model that displays a critical phase transition. It describes a miscibility phase transition on a lattice and has been solved exactly in two dimensions for the case of nearest neighbor interactions between components (17). It has been successfully used to describe critical behavior in a wide range of experimental systems very close to critical points with uncertainties of <10%, as detailed in Table 1. From the principle of universality, we expect the critical exponents characterizing the transition of our system to be identical to that of the two-dimensional Ising model because the thickness of our bilayer and the range of expected electrostatic interactions is much smaller than the correlation lengths we measure (18).

In our observation of critical phenomena, we benefit from distinct experimental advantages of employing a lipid bilayer. Correlation lengths occur on the scale of microns in our experiment, which permits us to use fluorescence microscopy to probe critical phenomena. These large correlation lengths derive from two factors. First, the critical temperature $T_c$ occurs near room temperature so that a reduced temperature $(T-T_c)/T_c$ on the order of $10^{-3}$ can be easily accessed with modest temperature control within 0.05°C. Second, the amplitude of correlation lengths in lipid membranes is relatively large, even when membranes are not near a critical point (19). Since our observed micron-scale correlation lengths are so much larger than the nanometer-scale thickness of a lipid membrane, the spectrum of critical fluctuations are those in the plane of the membrane and therefore are characteristic of the universality class of the *two-dimensional* Ising model. A benefit of probing fluctuations in two dimensions is that the Ginzburg criterion predicts a wider temperature region for critical behavior than in three dimensions (20).

We encounter two other experimental advantages of investigating critical behavior of liquid phases in lipid bilayers by fluorescence microscopy. First, our use of visible wavelengths enforces a resolution limit on the order of 0.5 microns. This length scale is significantly larger than the size of lipids (nanometers) or the range of interactions between lipids in a bilayer. Therefore, the fluorescence signal encoded in each pixel of our detector records the behavior of a very large number of lipids exhibiting critical behavior. In other words, in the critical regime, our images capture an ensemble of systems, each of which is itself behaving critically. Second, any impurities in our membrane are mobile, and do not constrain the scale over which critical fluctuations occur. In contrast, in systems of monolayers of atoms on solid surfaces, heterogeneities can limit the spatial range of fluctuations to a few hundred nanometers (e.g. (21)).

Our work complements other observations of composition fluctuations above $T_c$ and of fluctuations in domain boundaries below $T_c$. Above $T_c$, fluctuations have been detected as line broadening in NMR spectra. Veatch et al. observed that fluctuations on a scale of 50 nm persist in DPPC/DOPC/cholesterol vesicles over temperatures as high as 5-10°C above the miscibility transition (16). Recent theoretical work describes how broadening of nuclear magnetic resonance lineshapes arises due to composition fluctuations in membranes near critical points (16; 22). Below $T_c$, fluctuations of domain boundaries between liquid phases have been evaluated in both monolayers and bilayers to yield line tensions (9; 10; 12-15; 23-25). Several other experiments evaluate line tensions between a liquid in contact with either a gas or a solid (26-31). In contrast to the experiments described above, here we simultaneously investigate critical behavior both above and below the critical temperature. We combine systematic measurements of line tension, correlation length and lipid composition distributions in order to extract both critical exponents $\nu$ and $\beta$.

## METHODS

**Fluorescence Microscopy of Vesicles**
Free-floating giant unilamellar vesicles were made by electroformation (32; 33) from a lipid mixture of 25:20:55 mol% of diPhyPC, DPPC (Avanti Polar Lipids, Alabaster, AL) and cholesterol (Sigma, St. Louis, MO). We previously determined the phase diagram for this ternary system and identified this specific lipid ratio as being near a room temperature critical point. Membranes were labeled with 0.8 mol% Texas Red-DPPE (Invitrogen, Carlsbad, CA), which preferentially partitions into the liquid-disordered phase. Vesicles were used within four hours of electroformation. Consequently, the vesicles were taut with no domains bulging either inward or outward.

Vesicles were imaged with a frame acquisition time of 300ms using a 40x objective on a Nikon Y-25 microscope (Nikon, Melville, NY) coupled to a Coolsnap HQ CCD camera (Photometrics, Tucson, AZ). Temperature was maintained at the sample using a temperature controlled home-built copper microscope stage, which is the only heated element in our system. The microscope stage was enclosed in a Plexiglas box to eliminate drafts. Temperature was controlled using a Peltier thermoelectric device (Advanced Thermoelectric, Nashua, NH) operated by a current controller. The temperature sensor is a 10 kΩ thermistor read by the current controller with an optimum absolute accuracy quoted by the manufacturer as 0.02°C (both from Wavelength Electronics, Bozeman, MT). Our experiment measures a change in temperature, $T - T_c$, for which the measurement error should be smaller. The thermistor and sample lie side by side on identical 25mm x 25mm cover slips connected to the microscope stage by a thin layer of thermal grease (Omega Engineering, Stanford, CT). The thermistor is attached to its slide by conducting epoxy (Arctic Silver, Visalia, CA). The maximum temperature fluctuation measured over the course of an experiment was ±0.05°C. Usual fluctuations were measured to be ±0.01°C.

In all cases, video recordings of vesicles were begun above the miscibility transition temperature and continued as the temperature was stepped down until phase separation was complete. Vesicles were equilibrated in the dark for two minutes between each temperature step. Two minutes was ample time for vesicles to come to equilibrium as determined by the following experiment. A vesicle sample was equilibrated for two minutes after a temperature step, and correlation lengths were measured. The shutter was closed and vesicles remained in the dark at the same temperature for twenty minutes. Correlation lengths were again measured, and yielded the same values within error (data not shown). Movies were recorded over a series of temperatures for seven different vesicles. New vesicle batches were used for each series of video recordings.

**Analysis Overview**

For each vesicle, two independent analyses were conducted using custom software written in Matlab (The Mathworks, Natick, MA). Below the critical temperature, two phases were present and domain line tensions were obtained. Above the critical temperature, concentration fluctuations were visible, and correlation lengths of those fluctuations were determined. Vesicles were analyzed using both methods over the temperature range spanning 0.4-0.6°C of the anticipated critical temperature, in order to avoid bias in the assignment of $T_c$. We performed a second analysis above the critical temperature to extract the critical exponent $\beta$ by measuring the temperature dependence of histograms of pixel intensities of vesicle images.

**Analysis of Line Tension**

At low temperatures, images of the vesicle surface displayed distinct phase-separated domains. Our software traced domain perimeters through a series of movie frames. Each domain was analyzed independently. At the top left in figure 2, yellow lines trace perimeters as they fluctuate over about seven sec (about twenty frames), and are averaged to yield the red lines. A Fourier transform was performed on radial deviations $h(x)$ from the mean perimeter (as in the top right panel in figure 2) to yield reciprocal space deviations $h(k)$, where $k$ is the fluctuation wavenumber, with units of inverse length.

The capillary wave approximation for the energy of interface fluctuations is

$$E = \frac{1}{2}\lambda \int_0^L dx \left(\frac{\partial h}{\partial x}\right)^2 = \frac{\lambda L}{2}\sum_k k^2 h(k)h(-k), \qquad [1]$$

where $L$ is the smooth, mean perimeter of a domain, $k = 2\pi/L$, and $h(k)$ is the Fourier transform of $h(x)$:

$$h(k) = \frac{1}{L}\int_0^L dx\, h(x) e^{-ikx} . \qquad [2]$$

Since equation 1 is a quadratic form, the equipartition of energy theorem applies. Each capillary mode carries on average $k_B T/2$ in energy so that

$$\langle h(k)h(-k)\rangle = \langle |h(k)|^2\rangle = \frac{k_B T}{\lambda L k^2} , \qquad [3]$$

where the brackets denote an ensemble average, and $k_B$ is Boltzmann's constant (34; 35). The value of $\langle h(k)h(-k)\rangle$ is evaluated by averaging the value of $|h(k)|^2$ over all movie frames at one temperature. In figure 2, a log-log plot of $\langle h(k)h(-k)\rangle$ vs. $k$ confirms that the former scales as $k^2$. Figure 2 shows modes three through nine. The first two modes, which may not equilibrate fully over the course of a video, are not fit and are not shown in figure 2. We fit our log-log plot of $\langle h(k)h(-k)\rangle$ vs. $k$ and find the line tension from the intercept. All line tensions from one vesicle are collected in figure 3. Standard deviations are found by comparing line tensions from two or more fluctuating domains visible in a single frame and/or breaking long movies into runs of twenty frames or greater.

We are able to evaluate line tensions by applying the approximation of equation 1 to our data over a range of temperatures spanning about 10˚C below $T_c$ for two reasons. First, the correlation length is large enough that our observations are coarse-grained, so that the observed line tension is indeed the free energy per unit length. Second, the correlation length, while large, is still much smaller than the size of domains, which appear almost circular. All domains analyzed fell between 7.2 and 27 microns in diameter. Therefore the displacements of the domain boundaries are sufficiently small that the capillary wave approximation of equation 1 is valid. As long as it is applicable, displacements of the domain boundary on very small length scales do not contribute appreciably since the amplitude of each mode decays as $1/k^2$. Note that this method of extracting the critical exponent $\nu$ from the vanishing of the line tension $\lambda \sim (T_c - T)^\nu$ does not require the correlation length to be extremely large, i.e. for us to approach the critical point within a reduced temperature of $10^{-3}$. In fact, we reach the upper range of valid temperatures when correlation lengths are so large that domains no longer appear circular, but are now fractal as in figure 1 at 32.2˚C. The capillary wave description of these displacements is then no longer adequate. We reach the lower range of temperatures for which our measurement is valid when line tensions exceed about 1 pN, which results in perimeter fluctuations smaller than our optical resolution of approximately 0.5 µm at all values of $k$.

**Analysis of Correlation Length**

The grey scale intensity of each pixel in an image reports the average membrane composition inside that pixel. That average results from an integration over all fluctuations smaller than the pixel size. Fluctuations around the average lipid composition in the membrane are quantified by the two-point correlation function of an image where $I(\vec{r})$ is the image intensity (a density) at position $\vec{r}$:

$$G(\vec{r}) = \langle I(\vec{r}_0)I(\vec{r}_0 + \vec{r})\rangle - \langle I(\vec{r}_0)\rangle\langle I(\vec{r}_0 + \vec{r})\rangle \qquad [4]$$

The average, denoted by brackets, is defined to be over all overlapping pairs. We obtained the background corrected image by averaging grayscale values over all movie frames and subtracting the resulting value from each pixel of each movie frame.

Even though we observe critical fluctuations in real space, we found it convenient to perform our analysis in reciprocal space by obtaining the structure function $S(\vec{k})$ from a discrete Fourier transform of each image. The structure function is related to the correlation function by Fourier transform:

$$S(\vec{k}) = \langle I(\vec{k})I(-\vec{k})\rangle = \langle |I(\vec{k})|^2 \rangle = \frac{1}{(2\pi)^2}\int d\vec{r}\, G(\vec{r})e^{i\vec{k}\cdot\vec{r}}. \qquad [5]$$

Near $T_c$, the structure function is rotationally invariant, so we perform an angular average to obtain $S(k)$, which we then average over each frame of the movie. Because the pixel size of our camera (0.18 µm) is smaller than the optical resolution of our microscope (about 0.5 µm), our acquired images appear as if they are filtered by the point-spread function of the microscope, which has a full width at half maximum of 3.4 pixels, corresponding to 0.76 microns. We correct for the finite point-spread function of the microscope by dividing the evaluated structure factor in equation 5 by the structure factor of the point-spread function. This function was obtained by averaging acquired images of 20 individual 0.1 µm fluorescent beads (Invitrogen, Carlsbad CA). In our analysis, we measure intensity. Intensity is proportional, not equal, to the order parameter, *m*. As a consequence, we report *S(k)* in arbitrary units.

We image only one face of a vesicle rather than the whole vesicle. In general, a finite sized data set should result in a Fourier transform of the structure function that does not return the full two-point correlation function, but rather the product of that function and a window function reflecting the limited area imaged. However, our data are not sensitive to this problem because the correlation length that we measure is less than ten pixels, which is much smaller than the image size of 150-200 pixels in width. We use a discrete Fourier transform in Matlab which automatically imposes periodic boundary conditions. In general, periodic boundary conditions introduce extraneous correlations between points that span the boundary. We avoid this by supplementing the non-zero $I(\vec{r})$ with an equal number of zero values. Our data are also not sensitive to inhomogeneous light levels over the vesicle area that we sample, caused by the presence of other vesicles outside the focal plane. Removing long-wavelength gradients by filtering the image did not noticeably affect our measured correlation lengths (data not shown), so we present our data without this filtering.

The structure function $S(k)$ and the correlation function $G(r)$ for the two-dimensional Ising model are known exactly (36; 37). They do not follow simple analytic expressions, but are easily computed using the tabulated values of the scaling functions for $G(r)$ (36). If the transition in our system is in the universality class of the two-dimensional Ising model, then all our measured structure functions should agree with the scaling form $k^{7/4} S(k)$ vs. $k\xi$. In figure 4, the solid curve plotted on this scaling form is the Fourier transform of the exact numerical solution for $G(r)$ of the two-dimensional Ising model calculated by Wu et al. (36). In figure 4 we test whether our data agree with this scaling form and determine the correlation length, $\xi$, at each temperature by finding the value of $\xi$ by hand for which the data fall on the scaling form. We state uncertainty in $\xi$ as the range of values outside of which agreement with the exact scaling form is poor by eye, and it is an overestimate.

The fact that we can vary the single parameter $\xi$ to fit the function $k^{7/4} S(k)$ vs. $k\xi$ over a range of $k\xi$ for the eight temperatures shown in figure 4 is strong evidence that the transition is indeed in the universality class of the two-dimensional Ising model. In contrast to our analysis of the line tension, which does not require extreme proximity to the transition temperature, our method of fitting the scaling form performs best when the product $k\xi$ spans a wide range, in particular when the critical temperature is closely approached so that $\xi$ is large and $k\xi$ is of the order of unity.

To verify that our method of finding correlation lengths was appropriate, we performed a Monte Carlo simulation of a two-dimensional Ising model and extracted correlation lengths from the output images by the same methods as for the vesicle data. The simulation was implemented in Python (www.python.org) using the NumPy (www.scipy.org) and the Python Imaging Library (www.pythonware.com/products/pil/) modules. The simulation size was 200 x 200 sites and the simulation gave the expected value of $\nu = 1$. In the simulation, statistical fluctuating objects are single spins and the simulation is carried out at a reduced temperature $t_r = (T-T_c)/T_c$. In order to compare images produced by the simulation with images produced by the experiment in which each pixel contains $b$ x $b$ lipids, each fluctuating object in the simulation must represent a collection of $b^2$ spins. When $\nu = 1$, the scaling relation

$$\xi(t_r) = \frac{1}{b}\xi(b^{-1/\nu}t_r)$$

indicates that a snapshot of a simulation performed at $t_r/b$ in which each object contains $b^2$ spins is indistinguishable from a snapshot of a simulation performed at $t_r$ in which each object contains only one spin. Therefore, in figure 1 we quote reduced temperatures in the simulation as $t_r/b$ where $b = 225$, which is roughly the pixel size (224.75 nm) divided by an order of magnitude estimate for the inter-lipid distance (1nm). Correlation lengths derived from the simulation and the experiment are directly comparable if the reduced temperature is rescaled in this way.

One limitation that arises as the temperature approaches $T_c$ is that the dynamics of the system become very slow (20). Consequently brief movies (about seven seconds) do not consistently sample a full set of equilibrium configurations. We used quantitative structure factor data only if every visible composition fluctuation failed to persist over the length of one movie, to ensure that our system reached equilibrium. For each vesicle, movies were collected for nine to twelve temperatures above $T_c$. The two movies at temperatures nearest to $T_c$ were typically eliminated by the criterion above. Similar slow dynamics that occur below $T_c$ prevent the analysis of structure factors to find the critical exponent $\nu$ below $T_c$. We have tested that all movies used to find $\nu$ sufficiently sample the equilibrium state as follows. We altered our camera acquisition time by binning images from our videos in groups of two, increasing our effective acquisition time from 300 ms to 600 ms. Binning slightly changed the absolute values of the correlation lengths, but did not affect the critical exponent $\nu$. A different limitation arises at high temperatures above $T_c$. The smallest detectable correlation length in our system is related to our microscope's spatial resolution, and to the diminishing contrast between composition fluctuations as temperature increases.

**Analysis of pixel intensity distribution**

To determine the order parameter critical exponent $\beta$, we measured the temperature evolution of the distribution of pixel gray scales. The intensity of each pixel reports the composition of the membrane inside it via the fluorescent dye distribution. The probability distribution of the membrane's composition, or order parameter, has two peaks below $T_c$, separated by $\Delta m$. The distance between the double peaks decreases as $T_c$ is approached from below according to

$$\Delta m \propto \left| \frac{T - T_c}{T_c} \right|^{\beta} . \qquad [6]$$

Above $T_c$, there is a single peak at $m = 0$ whose width diverges as the critical temperature is approached. We extract the critical exponent $\beta$ by measuring the root-mean-square width of the single peak, which diverges as (38),

$$width = \langle m^2 \rangle^{1/2} \propto \left| \frac{T - T_c}{T_c} \right|^{-\beta} . \qquad [7]$$

An advantage of the procedure above is that we evaluate the width of the intensity distribution and can disregard the location of the center of the peak, which may drift to lower intensity values over the course of an experiment due to gradual photobleaching. In other words, our analysis of the distribution, like our analysis of line tension, does not depend on our greyscale calibration as $S(k)$ does. Moreover, our analysis of distribution widths, like our analysis of the line tension, does not rely explicitly on our system exhibiting very large correlation lengths compared to the pixel size, and therefore does not require the system to reach temperatures very close to the critical temperature. We have verified that our measurements are robust over the entire area of the vesicle that we sample. Specifically, we find no difference in mean intensity or noise between pixels from the center of our vesicle images and from the edge of the in-focus area that we sample. This result implies that our vesicles are so large that we can neglect their curvature over the area we sample. We have also verified that our measurement of $\beta$ does not depend on our camera acquisition time by binning images from our videos in groups of two, increasing our effective acquisition time from 300ms to 600ms. Binning changed the absolute values of the peak widths, but did not affect the critical exponent $\beta$. In principle, our procedure to obtain $\beta$ could be applied at temperatures below $T_c$. However, in our system, our inability to image the entire vesicle surface and our observation of slow dynamics prevents our video recordings from effectively sampling both peaks below $T_c$.

**Analysis of Critical Temperature and Exponent $\nu$**

As noted earlier, the line tension was measured below $T_c$ and vanishes as $T_c$ is approached like $\lambda \sim (T_c - T)^\nu$, whereas the correlation length was measured above $T_c$ and diverges as $\xi \sim |T_c - T|^{-\nu}$. The product of $\lambda$ and $\xi$ is a constant that depends on the definition of $\xi$ (39). Here we use the definition of $\xi$ within the two point correlation function $G(r)$, which has the following asymptotic form for large values of $r/\xi$ in all systems in the two-dimensional Ising universality class (36):

$$G(r) \sim \left(\frac{1}{\xi}\right)^{\frac{1}{4}} \left(\frac{\xi}{r}\right)^{\frac{1}{2}} \exp\left(\frac{-r}{\xi}\right) \qquad T > T_c \qquad [8]$$

$$G(r) - m^2 \sim \left(\frac{1}{\xi}\right)^{\frac{1}{4}} \left(\frac{\xi}{r}\right)^{2} \exp\left(\frac{-2r}{\xi}\right) \qquad T < T_c. \qquad [9]$$

A comparison of the scaling behavior of $G(r)$ as defined by Wu et al. (36) for $T > T_c$ and the line tension as given by Onsager (17) yields

$$\lambda \xi = k_B T_c \qquad [10]$$

within the scaling regime (37). We use equation 10 to compare values of the correlation length obtained from fits to the structure function above $T_c$ with those of line tension measured below $T_c$.

We find that plots of $k_B T / \xi$ above $T_c$ and of line tension below $T_c$ have roughly linear slopes, and that they intercept the $x$-axis at roughly the same temperature, as shown in the upper panel of figure 5. This temperature serves as our initial guess of $T_c$. In order to determine the exponent $\nu$ quantitatively, we produced log-log plots of $k_B T / \lambda$ below $T_c$ vs. the reduced temperature $(T - T_c)/T_c$, and of $\xi$ above $T_c$ vs. reduced temperature, as shown in the lower panel of figure 5. We assumed that the exponent $\nu$ is the same both above and below $T_c$. Therefore, we adjusted the value of $T_c$ until we found a single best $\nu$ that fit all data above and below $T_c$ according to the power law behavior

$$\xi = A \left| \frac{T - T_c}{T_c} \right|^{-\nu} \qquad [11]$$

where A is a constant amplitude (20).

**Photo-oxidation**

We made considerable efforts to reduce the adverse effects of photo-oxidation and are confident that photo-oxidation does not affect our results for the critical exponents outside our specified error bounds. It has been demonstrated previously that illumination of vesicles containing unsaturated phospholipids can alter the lipid composition of membranes and therefore modulate miscibility transition temperatures (40). This property has been used to control line tension of domains in phase separated vesicles (13). Our choice to work with membranes containing only saturated phospholipids significantly reduces the rate at which transition temperatures vary with light exposure. Furthermore, we used neutral density filters to reduce light intensity, and illumination of vesicles was limited to less than nine seconds at each temperature. These efforts reduce our measured drift in transition temperature due to photo-oxidation to at most -0.09°C per exposure. The effect of a gradual decrease in transition temperature would be to cause the critical exponent $\nu$ to appear smaller than its true value above $T_c$, and larger below. We performed a control experiment to ensure that photo-oxidation did not affect our measurements of both line tension and correlation length outside our specified error bounds. For vesicle C from figure 6, at 29.5°C line tension is $\lambda = 0.42 \pm 0.07$ pN over the course of the video. Over the first twenty frames of the video, before significant illumination, $\lambda = 0.37 \pm 0.16$ pN. Over the last twenty frames, after illumination, $\lambda = 0.47 \pm 0.40$ pN. Therefore, during this exposure, we observe no systematic decrease in either $T_c$ or $\lambda$, which would result from photo-oxidation. Our conclusion is that any photo-oxidation associated with illumination of the vesicle over the course of a video is minor such that line tension remains within our stated error bounds, which are standard deviations from multiple domains and/or movies. Similarly, for the same vesicle at 32.6°C, the correlation length derived from all available frames is $\xi = 1.17 \pm 0.54$ µm. The first five frames yield $\xi = 0.72$ µm and the last five yield $\xi = 0.69$ µm, both within our stated error.

**RESULTS**

We produce giant unilamellar vesicles whose membranes contain circular, liquid domains at low temperature (figure 1). At high temperature, the vesicle membranes are uniform. Vesicles were imaged over a series of temperatures, and images were analyzed in order to determine line tensions at low temperatures and correlation lengths at high temperatures. These results were then used to determine the critical exponents as described below.

**Line tension, $\lambda$**

Figure 2 shows that as the temperature approaches $T_c$, from below, fluctuations increase and line tension decreases. At 23.0°C, a temperature far below $T_c$, domains are circular, and their edges are smooth. At 31.5°C, which is close to the critical temperature, domains are oval, and their edges are irregular. We quantify the roughness of each domain boundary by compiling a spectrum of deviations of the domain radius from the mean domain radius, as described in Materials and Methods. Each spectrum yields a line tension. In the example in figure 2, we find $\lambda = 0.213 \pm 0.14$ pN for the spectrum at 23.0°C, and $\lambda = 0.019 \pm 0.012$ pN for the spectrum at 31.5°C.

The upper and lower panel of figure 3 show line tension measurements for all domains at all temperatures of two different vesicles. We find that line tension decreases linearly to zero at the critical temperature. Our measured line tension values are valid only for temperatures within 10°C of the critical temperature due to the optical resolution limit, as discussed in Materials and Methods.

**Correlation length, $\xi$**

The fluorescence micrographs in figure 1 show that fluctuations are small at temperatures far above $T_c$ and increase in size as the critical temperature is approached from above. Fluctuations span several microns very close to the critical temperature. This same observation is reflected in the measured structure factor of vesicle images, as shown for one vesicle in figure 4. At temperatures very close to $T_c$, structure factors have larger amplitudes and persist to larger wavenumbers.

Our structure factor data allow us to test whether we probe temperatures close enough to $T_c$ to capture behavior typical of the two-dimensional Ising universality class. We successfully adjusted the value of the correlation length, $\xi$, so that all structure factors for each vesicle superimposed on the exact form predicted by the two-dimensional Ising model (36) when plotted in the scaling form $k^{(7/4)} S(k)$ vs. $k\xi$, as shown in the bottom panel of figure 4.

**Critical exponent $\nu$**

Both line tension, $\lambda$, and energy per correlation length ($k_B T/\xi$) decrease roughly linearly to zero at the critical temperature. Figure 5 demonstrates how the critical temperature derived from line tension is within error of the critical temperature derived from energy per correlation length. These results are robust; figure 6 shows similar data for four additional vesicles. Each vesicle has a slightly different lipid composition, so each has a slightly different critical temperature.

Once we have determined the line tension below the critical temperature and the correlation length above it, we can extract the critical exponent $\nu$, as detailed in Methods. We require that $\nu$ have one unique value that fits data both above and below a single $T_c$ for a given vesicle. We obtain five values of $\nu$ from the five vesicles in figures 5 and 6, collected in table 2. The average of these five values, $\nu_{avg} = 1.2 \pm 0.2$, is consistent with that for the two-dimensional Ising model, $\nu = 1$ (17; 20).

**Critical exponent $\beta$**

Membrane composition, as reported by the intensity of fluorescent dye, is distributed around one value above $T_c$ and around two values below $T_c$. For example, in the top panel of figure 7, pixel intensity values are symmetrically distributed around a single value at $T = 34.0°C$, which is far above $T_c = 32.5 \pm 0.15°C$ for the particular vesicle analyzed. The distribution develops an asymmetry as temperature is lowered to $T = 32.6°C$, indicating the onset of spontaneous symmetry breaking as the membrane approaches $T_c$. At $32.4°C$, below $T_c$, the distribution is bimodal, and also asymmetric, possibly indicating that the movie was not long enough to sample both phases equally.

Plotting the width of all of the symmetric histograms for the vesicle in figure 7 provides an independent test of whether our data is described well by the two-dimensional Ising model. Equation 7 predicts that the width of the distribution is proportional to $(T - T_c)^{-\beta}$. In the lower panel of figure 7, we plot the distribution width as $(width)^{-8}$ vs. $T$. Our data can be fit to a straight line, in agreement with the two-dimensional Ising prediction of $\beta = 1/8$. It is also reassuring that the critical temperature extracted from figure 7 is $\sim 32.8°C$, close to the value of $T_c \sim 32.5 \pm 0.15°C$ determined previously.

**DISCUSSION**

We have measured the line tension below the critical point and both the structure function and the distribution of the membrane composition above the critical point. We have extracted critical exponents of $\nu = 1.2 \pm 0.2$ and $\beta = 0.124 \pm 0.03$ from our data. These values are in agreement with those of the universality class of the two-dimensional Ising model, $\nu = 1$ and $\beta = 0.125$. The accord between our experimental results and theoretical predictions is also clearly seen in the collapse of experimental structure factors onto the exact result from the two-dimensional Ising model, in figure 4. This is the first quantitative measurement of critical exponents in a lipid bilayer with liquid phases.

Agreement with the theory is excellent despite our experimental limitations in spatial resolution, in proximity to the critical temperature, and in variation of lipid compositions between vesicles. Vesicles made by electroformation generally contain a variety of compositions distributed around the bulk composition, and as a consequence, we observe a broad range of transition temperatures in this mixture, consistent with earlier work (33), meaning that small changes in composition result in relatively large differences in miscibility transition temperature. Despite the inherent distribution in vesicle composition in our system, we observe critical behavior in all vesicles probed, meaning that they are well within the critical region. For two vesicles #3 and #4 in table 2, we find values of the critical exponent $\nu$ that are significantly greater than the predicted value of $\nu = 1$. One possible explanation for this deviation is that these vesicles have slightly off-critical compositions.

In addition to critical exponents, we have also determined the correlation length above the critical temperature. We have compared the thermal energy per correlation length ($k_BT/\xi$) above $T_c$ with the line tension below $T_c$. These two quantities are predicted by the two-dimensional Ising model and universality to be equal in the scaling regime (37). Figures 5 and 6 show that the two quantities are in reasonable agreement given our limitations in obtaining large correlations needed for analysis of structure functions.

Below $T_c$, we measure line tensions from the maximum measurable value of ~0.8 pN to a value of ~0.05 pN near the critical point. Our highest measured line tensions are consistent with estimates of line tension between 0.5 and 3.3 pN using flaccid vesicles composed of a similar lipid mixture (10; 12). The flaccid vesicle method requires that line tension is large compared with the membrane bending modulus (12). The measurement method we present here is complementary because it requires that line tension is low for fluctuations to be visible (10). Recent work observed line tensions between domains as small as 0.05 pN in vesicles passing through critical points (13). Instead of approaching the critical point by changing temperature, as in our work, the researchers changed the vesicle's composition by photo-oxidizing lipids (13).

## SPECULATIONS ON BIOLOGICAL SIGNIFICANCE

We have shown that in lipid vesicles, fluctuations large enough to observe by light microscopy (greater than 1 micron) appear within 0.5°C of $T_c$. Extrapolating from our data using equation 8, with $\nu = 1$, we expect sub-micron fluctuations with correlation lengths of ~50 nm to occur in vesicles between 2-8°C above their critical temperature. This prediction is quantitatively consistent with recent reports of ~50nm fluctuations in near-critical membranes of DOPC/DPPC/cholesterol probed by NMR (16).

In plasma membranes of unstimulated cells, no micron-scale domains are observed by fluorescence microscopy at the cells' growth temperature. Therefore, domains or composition fluctuations must be sub-micron in dimension if they are present. Sub-micron, local differences in membrane composition may confer advantages for cell processes (41). Although the concept currently in favor is that sub-micron membrane heterogeneity in cell membranes is driven by phase separation, there is no consensus on what mechanism might prevent sub-micron domains from ripening into larger domains (42). It has been suggested that dynamic, small-scale membrane heterogeneities could result from critical fluctuations near a critical temperature, rather than small domains far below $T_c$ that are prevented from coalescing (16; 31; 43; 44). Here we have shown that it is possible to tune domain size (and line tension) by changing the membrane's proximity to a miscibility critical point. It has been reported that vesicles isolated from the plasma membranes of living RBL mast cells and other cell types also display critical behavior of the type exhibited here (45).

## ACKNOWLEDGEMENTS

We thank David Burdick for assistance in running simulations and the lab of Phil Reid for the loan of 0.1 µm fluorescent beads. We thank Ben Widom, James Sethna, Ben Stottrup, Tobias Baumgart, and Peter Olmsted for helpful discussions. We also thank the anonymous reviewers for their careful readings of the manuscript. SLK acknowledges support from the National Science Foundation (MCB-0133484) and a Research Corporation Cottrell Scholar Award. ARHS was supported by a UW Center for Nanotechnology IGERT Award NSF #DGE-0504573 and a Molecular Biophysics Training grant NIH #5 T32 GM08268-20. MDC was supported by an NIH Ruth Kirschstein Postdoctoral Fellowship. SLV is supported through a postdoctoral fellowship from the Cancer Research Institute. MdN and MS acknowledge NSF grants DMR-0341341 and 0503752, respectively.

# REFERENCES

1. Silvius, J.R. 1982. Thermotropic phase transitions of pure lipids in model membranes and their modifications by membrane proteins. John Wiley, New York.
2. Lindsey, H., N.O. Petersen, and S.I. Chan. 1979. Physiochemical characterization of 1,2-diphytanoyl-sn-glycero-3-phosphocholine in model membrane systems. *Biochim. Biophys. Acta* 555:147-167.
3. Veatch, S.L., and S.L. Keller. 2003. Separation of liquid phases in giant vesicles of ternary mixtures of phospholipids and cholesterol. *Biophys. J.* 85:3074-3083.
4. Cicuta, P., S.L. Keller, and S.L. Veatch. 2007. Diffusion of liquid domains in lipid bilayer membranes. *J. Phys. Chem. B* 1111:3328-3331.
5. Bacia, K., D. Scherfeld, N. Kahya, and P. Schwille. 2004. Fluorescence correlation spectroscopy relates rafts in model and native membranes. *Biophys. J.* 87:1034-1043.
6. Filippov, A., G. Orädd, and G. Lindblom. 2006. Sphingomyelin structure influences the lateral diffusion and raft formation in lipid bilayers. *Biophys. J.* 90:2086-2092.
7. Ipsen, J.H., G. Karlström, O.G. Mouritsen, H. Wennerström, and M.J. Zuckermann. 1987. Phase equilibria in the phosphatidylcholine-cholesterol system. *Biochim. Biophys. Acta* 905:162-172.
8. Veatch, S.L., K. Gawrisch, and S.L. Keller. 2006. Closed-loop miscibility gap and quantitative tie-lines in ternary membranes containing diphytanoyl PC. *Biophys. J.* 90:4428-4436.
9. Benvegnu, D.J., and H.M. McConnell. 1992. Line tension between liquid domains in lipid monolayers. *J. Phys. Chem.* 96:6820-6824.
10. Baumgart, T., S.T. Hess, and W.W. Webb. 2003. Imaging coexisting fluid domains in biomembrane models coupling curvature and line tension. *Nature* 425:821-824.
11. Veatch, S.L., I.V. Polozov, K. Gawrisch, and S.L. Keller. 2004. Liquid domains in vesicles investigated by NMR and fluorescence microscopy. *Biophys. J.* 86:2910-2922.
12. Tian, A., C. Johnson, W. Wang, and T. Baumgart. 2007. Line tension at fluid membrane boundaries measured by micropipette aspiration. *Phys. Rev. Lett.* 98:208102.
13. Esposito, C., A. Tian, S. Melamed, C. Johnson, S.-Y. Tee, and T. Baumgart. 2007. Flicker spectroscopy of thermal lipid bilayer domain boundary fluctuations. *Biophys. J.* 93:3169-3181.
14. Stottrup, B.L., A.M. Heussler, and T.A. Bibelnieks. 2007. Determination of line tension in lipid monolayers by Fourier analysis of capillary waves. *J. Phys. Chem. B.* 111:11091-11094.
15. García-Sáez, A.J., S. Chiantia, and P. Schwille. 2007. Effect of line tension on the lateral organization of lipid membranes. *J. Biol. Chem.* 282:33537-33544.
16. Veatch, S.L., O. Soubias, S.L. Keller, and K. Gawrisch. 2007. Critical fluctuations in domain-forming lipid mixtures. *Proc. Natl. Acad. Sci. USA* 104:17650-17655.

17. Onsager, L. 1944. Crystal statistics. I. A two-dimensional model with an order-disorder transition. *Phys. Rev.* 65:117-149.
18. Sengers, J.V., and J.M.H. Levelt Sengers. 1986. Thermodynamic behavior of fluids near the critical point. *Annu. Rev. Phys. Chem.* 37:189-222.
19. Spaar, A., and T. Salditt. 2003. Short range order of hydrocarbon chains in fluid phospholipid bilayers studied by x-ray diffraction from highly oriented membranes. *Biophys. J.* 85:1576-1584.
20. Goldenfeld, N. 1992. Lectures on phase transitions and the renormalization group. Addison-Wesley, New York.
21. Tejwani, M.J., O. Ferriera, and O.E. Vilches. 1983. Possible Ising transition in a 4He monolayer absorbed on Kr-plated graphite. *Phys. Rev. Lett.* 44:152-155.
22. Radhakrishnan, A., and H. McConnell. 2007. Composition fluctuations, chemical exchange, and nuclear relaxation in membranes containing cholesterol. *J. Phys. Chem. B* 126:185101.
23. Goldstein, R.E., and D.P. Jackson. 1994. Domain shape relaxation and the spectrum of thermal fluctuations in Langmuir monolayers. *J. Phys. Chem.* 98:9626-9636.
24. Seul, M. 1990. Domain wall fluctuations and instabilities in monomolecular films. *Physica A* 168:198-209.
25. Mann, E.K., S. Hénon, D. Langevin, and J. Meunier. 1992. Molecular layers of a polymer at the free water surface: Microscopy at the Brewster angle. *J. Phys. II France* 2:1683-1704.
26. Muller, P., and F. Gallet. 1991. First measurement of the liquid-solid line energy. *Phys. Rev. Lett.* 67:1106-1109.
27. Riviere, S., S. Hénon, J. Meunier, G. Albrecht, M.M. Boissonnade, and A. Baszkin. 1995. Electrostatic pressure and line tension in a Langmuir monolayer. *Phys. Rev. Lett.* 75:2506-2509.
28. Riviere, S., S. Hénon, and J. Meunier. 1994. Fluctuations of a defect line of molecular orientation in a monolayer. *Phys. Rev. E* 49:1375-1382.
29. Mann, E.K., S. Hénon, D. Langevin, J. Meunier, and L. Léger. 1995. Hydrodynamics of domain relaxation in a polymer monolayer. *Phys. Rev. E* 51:5708-5720.
30. Wurlitzer, S., P. Steffen, and T.M. Fischer. 2000. Line tension of Langmuir monolayer phase boundaries determined with optical tweezers. *J. Chem. Phys.* 112:5915-5918.
31. Nielsen, L.K., T. Bjørnholm, and O.G. Mouritsen. 2007. Thermodynamic and real-space structural evidence of a 2D critical point in phospholipid monolayers. *Langmuir* 23:11684-11692.
32. Angelova, M.I., S. Soléau, P. Méléard, J.F. Faucon, and P. Bothorel. 1992. Preparation of giant vesicles by external AC electric fields. *Progr Colloid Polym Sci* 89:127-131.
33. Veatch, S.L., and S.L. Keller. 2005. Seeing spots: Complex phase behavior in simple membranes. *Biochim. Biophys. Acta* 1746:172-185.
34. Safran, S.A. 1994. Statistical thermodynamics of surfaces, interfaces, and membranes. Addison-Wesley, Reading, MA.

35. Aarts, D.G.A.L., M. Schmidt, and H.N.W. Lekkerkerker. 2004. Direct visual observation of thermal capillary waves. *Science* 304:847-850.
36. Wu, T.T., B.M. McCoy, C.A. Tracy, and E. Barouch. 1976. Spin-spin correlation functions for the two-dimensional Ising model: Exact theory in the scaling region. *Phys. Rev. B* 13:316-374.
37. Itzykson, C., and J.-M. Drouffe. 1992. Statistical Field Theory Vol. 1. Cambridge University Press, New York.
38. Binder, K. 1981. Finite size scaling analysis of Ising model block distribution functions. *Zeitschrift fur Physik B* 43:119-140.
39. Widom, B. 1965. Surface tension and molecular correlations near the critical point. *J. Chem. Phys.* 43:3892-3897.
40. Ayuyan, A.G., and F.S. Cohen. 2006. Lipid peroxides promote large rafts: Effects of excitation of probes in fluorescence microscopy and electrochemical reactions during vesicle formation. *Biophys. J.* 91:2172-2183.
41. Nicolau Jr., D.V., K. Burrage, R.G. Parton, and J.F. Hancock. 2006. Identifying optimal lipid raft characteristics required to promote nanoscale protein-protein interactions on the plasma membrane. *Mol. Cell. Biol.* 26:313-323.
42. Edidin, M. 2003. The state of lipid rafts: from model membranes to cells. *Annu. Rev. Biophys. Biomol. Struct.* 32:257-283.
43. Nielsen, L.K., T. Bjørnholm, and O.G. Mouritsen. 2000. Fluctuations caught in the act. *Nature* 404:352.
44. Veatch, S.L. 2007. From small fluctuations to large-scale phase separation: Lateral organization in model membranes containing cholesterol. *Semin. Cell Dev. Biol.* 18:573-582.
45. Veatch, S.L., P. Cicuta, P. Sengupta, A. Honerkamp-Smith, D. Holowka, and B. Baird. 2007. Critical fluctuations in plasma membrane vesicles. *submitted*.
46. Rowlinson, J.S., and B. Widom. 1982. The molecular theory of capillarity. Clarendon Press, Oxford.
47. Hagen, J.P., and H.M. McConnell. 1997. Liquid-liquid immiscibility in lipid monolayers. *Biochim. Biophys. Acta* 1329:7-11.
48. Samuelsen, E.J. 1974. Critical behaviour of the two-dimensional Ising antiferromagnets $K_2CoF_4$ and $Rb_2CoF_4$. *J. Phys. Chem. Solids* 35:785-793.
49. Campuzano, J.C., M.S. Foster, G. Jennings, R.F. Willis, and W. Unertl. 1985. Au(110) (1 x 2) -to- (1 x 1) phase transition: A physical realization of the two-dimensional Ising model. *Phys. Rev. Lett.* 54:2684-2687.
50. LaBella, V.P., D.W. Bullock, M. Anser, Z. Ding, C. Emery, L. Bellaiche, and P.M. Thibado. 2000. Microscopic view of two-dimensional lattice-gas Ising system within the grand canonical ensemble. *Phys. Rev. Lett.* 84:4152-4255.

Table 1. Two-dimensional critical exponents

| System | Equation | Exponent | Exact Values (46) | Ref. |
|---|---|---|---|---|
| Monolayers of phospholipids (DMoPC, DPoPc, DOPC or DEPE) and dihydrocholesterol with 0.25% TR-DPPE (dye) | dihydrocholesterol concentration $\|x-x_c\| \sim (\Pi_c - \Pi)^\beta$ | $\beta = 0.25$ $\pm 0.07$ | $\beta = 0.125$ | (47) |
| Ising-like crystals of $K_2CoF_4$ | magnetic order parameter $m \sim \|T-T_c\|^\beta$ | $\beta = 0.116$ $\pm 0.008$ | $\beta = 0.125$ | (48) |
| Ising-like crystals of $Rb_2CoF_4$ | magnetic order parameter $m \sim \|T-T_c\|^\beta$ | $\beta = 0.119$ $\pm 0.008$ | $\beta = 0.125$ | (48) |
| " | inverse correlation length $K \sim (T - T_c)^\nu$ | $\nu = 0.89$ $\pm 0.1$ | $\nu = 1$ | |
| (110) facet reconstruction of Au surface | order parameter $m \sim \|T-T_c\|^\beta$ | $\beta = 0.13$ $\pm 0.022$ | $\beta = 0.125$ | (49) |
| " | correlation length $\xi \sim \|T-T_c\|^{-\nu}$ | $\nu = 1.02$ $\pm 0.02$ | $\nu = 1$ | |
| GaAs crystal surfaces | correlation length $\xi \sim \|T-T_c\|^{-\nu}$ | $\nu = 1.0$ $\pm 0.25$ | $\nu = 1$ | (50) |
| Lipid monolayer of 69/30/1 DMPC/cholesterol/NBD-PC (dye) | line tension $\lambda = m(\Pi_c - \Pi)^\nu$ | $\nu = 1.0 - 1.4$ | $\nu = 1$ | (9) |
| Helium on Kr-plated graphite | Heat capacity critical exponent, $\alpha$, via $2\nu = 2 - \alpha$ | $\nu = 0.86$ $\pm 0.05$ | $\nu = 1$ | (21) |

Table 2.  Values and standard deviations of the correlation length exponent, $\nu$, for vesicles in figures 5 and 6.

| Vesicle | Measured $\nu$ | S.D. in $\nu$ above $T_c$ | S.D. in $\nu$ below $T_c$ |
|---|---|---|---|
| 1 | 1.01 | ± 0.48 | ± 0.17 |
| 2 | 1.11 | ± 0.49 | ± 0.14 |
| 3 | 1.33 | ± 0.14 | ± 0.19 |
| 4 | 1.32 | ± 0.14 | ± 0.06 |
| 5 | 1.06 | ± 0.18 | ± 0.07 |

**FIGURE CAPTIONS**

**Figure 1.** Giant unilamellar vesicle passing through a miscibility critical point at $T_c$~32.5°C. Scale bar is 20 um. The bottom row (A-C) shows Ising model simulations at rescaled temperatures as described in the Methods section. The simulation in Panel A was performed at a reduced temperature comparable to the vesicle at 32.8°C, Panel B to the vesicle at 32.6°C, and Panel C to the vesicle at 32.4°C.

**Figure 2.** *Top left:* Frames from two movies of a giant unilamellar vesicle with a critical temperature of $T_c$ = 31.7 ±0.14°C. Domain boundaries fluctuate more at 31.5°C than at a temperature further from $T_c$ (23.0°C). Yellow lines trace all boundaries over a series of exposures taken over ~7s and red lines denote average boundaries. Scale bar is 20 μm. *Top right:* The line tension around any domain is found by first compiling a power spectrum of radial deviations ($h(x)$) of the boundary. *Bottom:* Log-log plot of the spectrum of radial deviations vs. inverse length $k$ for two individual domains, one with a mean perimeter of 86.7 μm at 23.0°C (triangles) and the other with a mean perimeter of 34.1 μm at 31.5°C (circles). Lines are fits to the equation $<h(k)^2> = k_BT/(L\lambda k^2)$. Only the data points from the third through the ninth wavenumbers are used for the fit. Error bars represent standard deviations from measurements over a series of movie frames.

**Figure 3.** The two panels show domain line tension with standard deviations measured for two different vesicles, at various temperatures. The line tension decreases linearly with increasing temperature close to $T_c$. Standard deviations are found by comparing line tensions from two or more fluctuating domains visible in a single frame and/or breaking long movies into runs of ≥20 frames. At some temperatures, only one domain was available and no standard deviation is shown.

**Figure 4**. *Top:* Measured structure factors are shown for one vesicle at eight temperatures. The lowest temperature shown is close to the critical temperature for this vesicle, $T_c$ = 26.43°C, derived by fitting critical exponents as described in the text. *Bottom:* When each structure factor is rescaled by a correlation length, $\xi$, all curves overlap with the exact form for the two-dimensional Ising model. The Ising model curve is the Fourier transform of the exact numerical solution for $G(r)$ calculated by Wu et al. (36).

**Figure 5.** *Top:* Line tension and structure factor data for the vesicle in figure 4 are combined in one panel. Filled circles denote measured line tensions, $\lambda$, measured below $T_c$. Error bars indicate standard deviations from three or more measurements. Filled triangles denote energy per correlation length, $k_BT/\xi$, measured above $T_c$. Uncertainties in correlation length are generated as described in Materials and Methods and are overestimates. Both data sets decrease linearly and reach zero at approximately the same temperature, 26.16 ± 0.16°C for line tension data and 26.52 ± 1.2°C for $k_BT/\xi$ data.

*Bottom:* When correlation length, $\xi$, or energy per line tension, $k_BT/\lambda$, is plotted on a log-log scale vs. reduced temperature $|T - T_c| / T_c$, the slope of the line yields the critical exponent, $\nu$. The solid lines show a good fit to $\nu = 1$, predicted by the two-dimensional Ising model. The dashed lines show a poor fit to $\nu = 1/2$, predicted by mean field theory, which would be applicable only if the system were not in the critical regime.

**Figure 6.** (A-D) The same procedures in Figures 1-5 are repeated for four additional vesicles. All together, the data show that line tension and energy per correlation length decrease approximately linearly to a minimum value of zero at the critical temperature, $T_c$. Each vesicle has a slightly different composition, and hence a slightly different $T_c$. Measured line tension, $\lambda$, is plotted below $T_c$ using filled circles. Energy per correlation length, $k_BT/\xi$, is plotted above $T_c$ using filled triangles.

**Figure 7.** *Top:* Histograms of pixel intensity (from 0 to 256) for movies of the same vesicle in Figure 1 are averaged over all frames at three temperatures. The distribution is peaked around one value at $T = 34.0°C$, which is far above the critical temperature of $T_c \approx 32.5 \pm 0.15°C$ for this system. At 32.6°C, the distribution is slightly asymmetric. Below $T_c$, at 32.4°C, the distribution appears bimodal, indicating $T < T_c$. It is also strongly asymmetric, indicating that the movie is not long enough to capture all states of the vesicle domains. The histogram from a vesicle containing one uniform phase is significantly narrower. For example, the solid line shows a histogram from a uniform vesicle of DOPC and 0.8% Texas Red-DPPE at 23˚C. *Bottom:* The second moment (*width*) of the averaged histograms is plotted in the form (*width*)$^{-8}$. The straight line shows that the data are consistent with an order parameter exponent $\beta = 1/8$, as predicted by the two-dimensional Ising model. The dashed line shows that the data is inconsistent with the mean field exponent of $\beta = 1/2$, which would be applicable only if the system were not in the critical regime. Only data from symmetrical peaks are included in the fit.

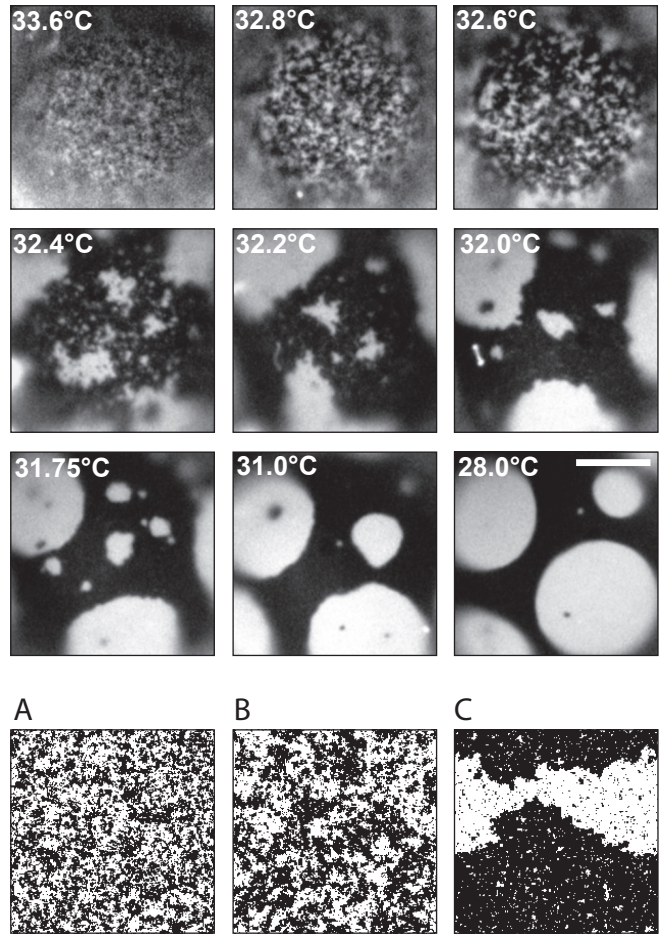

Figure 1

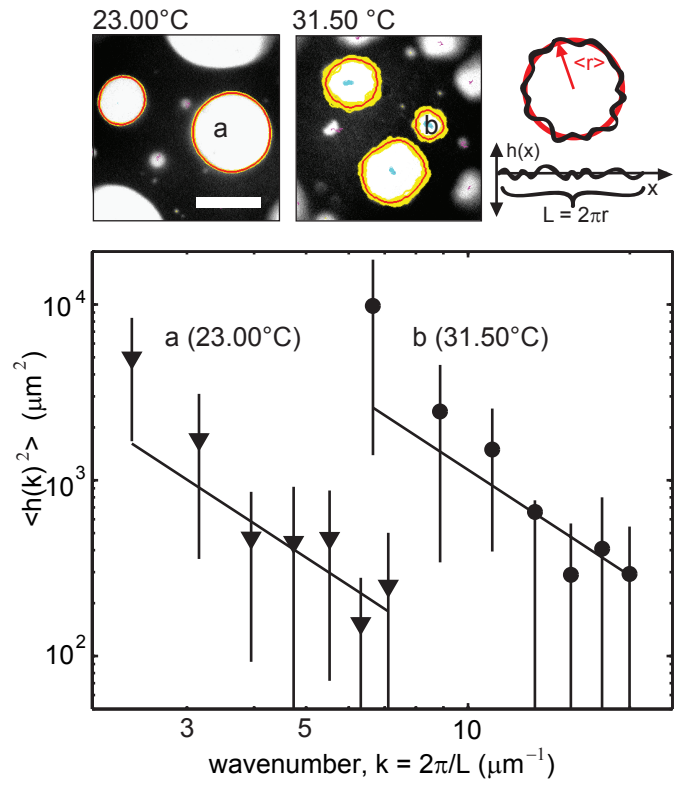

Figure 2

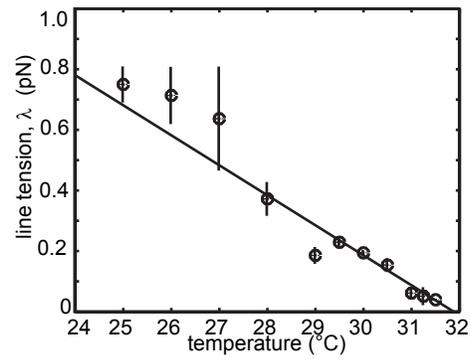
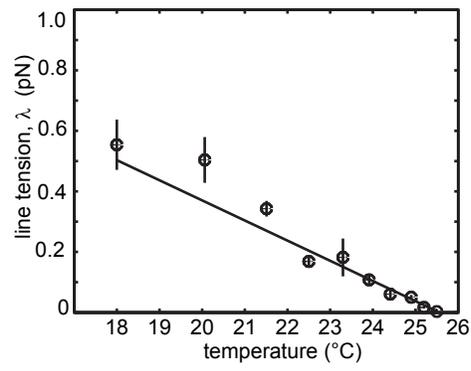

Figure 3

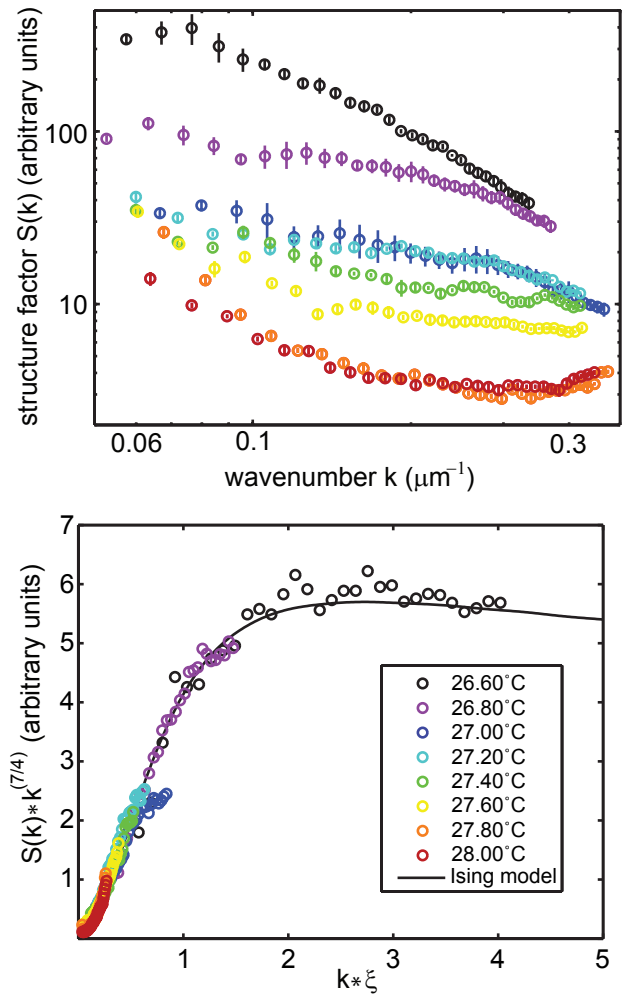

Figure 4

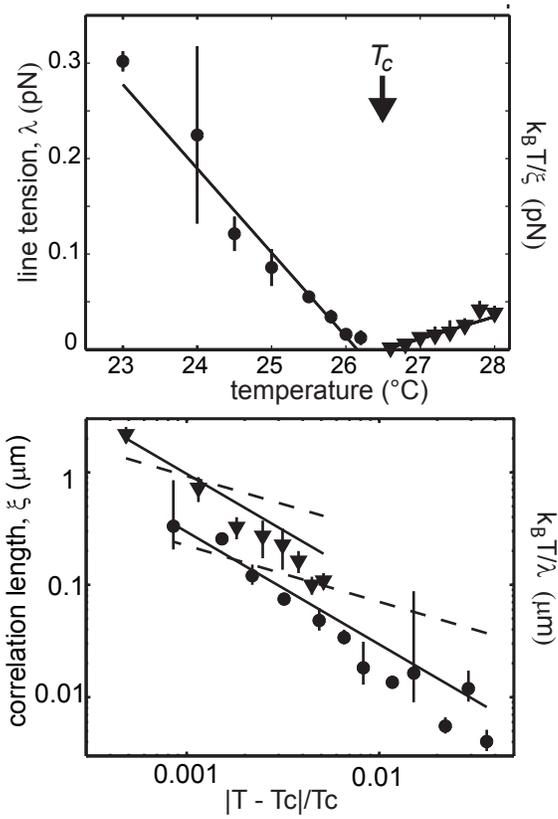

Figure 5

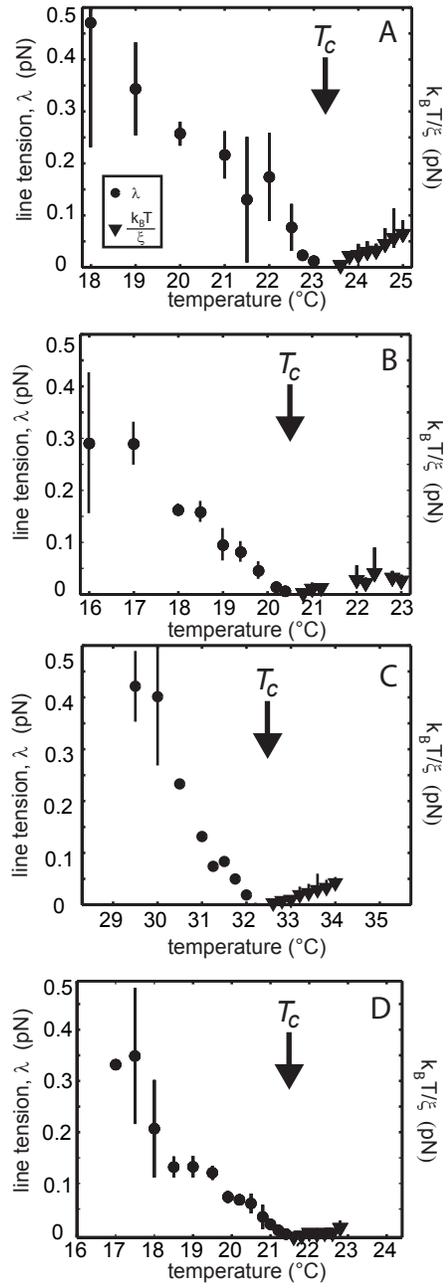

Figure 6

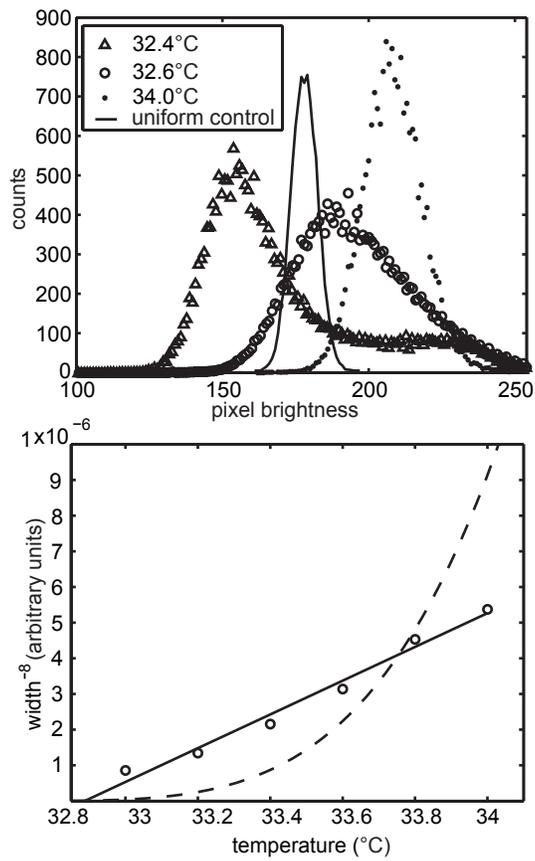

Figure 7